\documentclass[useAMS,usenatbib]{mn2e}
\usepackage{amssymb}
\usepackage{bm}
\usepackage{amsfonts}
\usepackage{amsmath}
\usepackage{graphicx}
\usepackage{subfigure}

\usepackage{epsf}

\newcommand{\etal}{et al.~}
\def\gsim{\lower 2pt \hbox{$\, \buildrel {\scriptstyle >}\over
{\scriptstyle \sim}\,$}}
\def\lsim{\lower 2pt \hbox{$\, \buildrel {\scriptstyle <}\over
{\scriptstyle \sim}\,$}}

\def\civ{C~{\scriptsize IV}}

\def\nv{N~{\sc v}}

\def\ovi{O~{\sc vi}}

\def\siiv{Si~{\scriptsize IV}}

\def\siii{Si~{\scriptsize II}}

\newcommand{\aap}{A\&A}
\newcommand{\nat}{Nature}
\newcommand{\pasp}{PASP}
\newcommand{\iaucirc}{IAUC}
\newcommand{\aj}{AJ}
\newcommand{\apj}{ApJ}
\newcommand{\apjl}{ApJ}
\newcommand{\apjs}{ApJS}
\newcommand{\araa}{ARA\&A}

\title[BH mass measurement from wind absorption lines]{Measuring the black hole masses in accreting X-ray binaries by detecting the Doppler orbital motion of their
accretion disk wind absorption lines}
\author[S. N. Zhang, J. Liao \& Y. Yao]{Shuang-Nan Zhang $^{1,\ 2}$, Jinyuan Liao $^{1}$, Yangsen Yao $^{3}$\thanks{E-mail:
zhangsn@ihep.ac.cn}\\
$^{1}$ Key Laboratory of Particle Astrophysics, Institute of High Energy Physics, Chinese Academy of Sciences, Beijing 100049, China\\
$^{2}$ Space Science Division, National Astronomical Observatories of
China, Chinese Academy of Sciences, Beijing 100012, China\\
$^{3}$ Center for Astrophysics and Space Astronomy, University of Colorado, Boulder, CO 80309, USA}

\begin{document}

\date{Accepted date. Received  date}

\pagerange{\pageref{firstpage}--\pageref{lastpage}} \pubyear{2011}

\maketitle

\label{firstpage}

\begin{abstract}

So far essentially all black hole masses in X-ray binaries have been obtained by observing the companion star's
velocity and light curves as functions of the orbital phase. However a major uncertainty is the estimate of the
orbital inclination angle of an X-ray binary. Here we suggest to measure the black hole mass in an X-ray binary
by measuring directly the black hole's orbital motion, thus obtaining the companion to black hole mass ratio. In
this method we assume that accretion disk wind moves with the black hole and thus the black hole's orbital
motion can be obtained from the Doppler velocity of the absorption lines produced in the accretion disk wind. We
validate this method by analyzing the Chandra/HETG observations of GRO~J1655--40, in which the black hole
orbital motion ($K_{\rm BH}=90.8\pm 11.3$~km s$^{-1}$) inferred from the Doppler velocity of disk-wind
absorption lines is consistent with the prediction from its previously measured system parameters. We thus
estimate its black hole mass ($M_{\rm BH}=5.41^{+0.98}_{-0.57}~M_{\odot}$) and then its system inclination
($i=72.0^{+7.8}_{-7.5}~^\circ$), where $M_{\rm BH}$ does not depend on $i$. Additional observations of this
source covering more orbital phases can improve estimates on its system parameters substantially. We then apply
the method to the black hole X-ray binary LMC~X--3 observed with HST/COS near orbital phase 0.75. We find that
the disk-wind absorption lines of \civ\ doublet were shifted to $\sim50~{\rm km~s^{-1}}$, which yields a
companion-to-black-hole mass ratio of 0.6 for an assumed disk wind velocity of $-400~{\rm km~s^{-1}}$.
Additional observations covering other orbital phases (0.25 in particular) are crucial to ease this assumption
and then to directly constrain the mass ratio. This method in principle can also be applied to any accreting
compact objects with detectable accretion disk wind absorption line features.
\end{abstract}

\section{Introduction}

Black holes (BHs) are believed to exist in many X-ray binaries (XRBs) and
active galactic nuclei (AGN). Measurement of the motion of a BH with
respect to its surrounding in such a system can in principle set strong
constraints on the mass of the BH. However in an AGN this is normally not
possible because the time scale of
BH's significant motion is too long and/or the BH is barely moving at
all. In an XRB, the BH should move with
respect to the system's center of mass (CM), making it possible to detect directly the BH's motion with respect
to the CM. However in a low-mass XRB (LMXB) the companion is normally much less massive than the BH, so that the
BH barely moves or moves very slowly with respect to the CM. In some LMXBs, the companion's mass is comparable
or only several times less massive than the BH, e.g., in GRO~J1655--40 (\citealt{1994IAUC.6106....1Z,
1997ApJ...477..876O}), the BHs' motion may be significant enough for direct detection. The most favorable
systems for detecting BH's motion should be high-mass XRBs (HMXBs), in which the BHs move rapidly with respect
to their CMs.

Orbital motion of double-peaked disk emission lines were observed for neutron star XRB Sco X--1
(\citealt{sc02}), the BH XRB A0620--00 (\citealt{1990ApJ...359L..47H,1994ApJ...436..848O}), and the BH GRS
1124--68 (\citealt{1994ApJ...436..848O}). Unfortunately a significant phase offset of velocity modulation was
found from that expected based on the observed orbital motion of the companion, though the velocity
semi-amplitude is consistent with the expected mass ratio (\citealt{1994ApJ...436..848O}). Soria et al. (1998)
observed the orbital motion of the double-peaked disk emission line He~{\scriptsize II}~$\lambda$4686 from
GRO~J1655--40, and found its velocity modulation phase and semi-amplitude in agreement with the kinematic and
dynamical parameters of the system. Therefore a more robust mass lower limit is placed based on the observed
motion of the primary and thus ruling out any possibility for a neutron star as the primary in the system (Soria
et al. 1998). However one major problem in accurately measuring the orbital motion of the primary from the
observed double-peaked emission lines is how to determine reliably the line center, because the lines are
typically asymmetric and also variable.

In an XRB, both the accretion disk and its wind move with the BH, and thus provides us with another opportunity
to measure the BH's motion via Doppler shift of absorption features of the accretion disk wind. Accretion disk
winds are ubiquitous in XRBs and normally detected through ionized absorption lines, typically with around
$\sim10^3 {\rm km~s^{-1}}$ or less (e.g., \citealt{ueda04,mil04,mil06a,mil08}),  but can reach to about
$\sim10^4 {\rm km~s^{-1}}$ in some extreme cases, e.g. in the newly discovered BH transient IGR~J17091--3624
(\citealt{2011arXiv1112.3648K}). In particular the high quality Chandra/HETG observations of the BHXB
GRO~J1655--40 have found many highly ionized narrow absorption lines, which are interpreted as evidence of
magnetic field-driven accretion disk wind (\citealt{mil06a,2009ApJ...701..865K,luk10}); although the absorption
lines are also interpreted as from the absorption by X-ray heated thermal wind (Netzer 2006). Regardless the
origin of the accretion disk wind, the orbital motion of its many absorption lines may be measured reliably,
because typically many narrow absorption lines are present with high signal to noise ratios and appear to be
rather stable when observed.

In this work, we model the Doppler motion of wind absorption lines from the
LMXB GRO~J1655--40 and HMXB LMC~X--3
to constrain directly the companion to primary mass ratio, in order to measure their BH masses and orbital
inclination angles. We first describe the methodology and test its
feasibility by applying it to Chandra/HETG observations of GRO~J1655--40,
revealing for the first time the velocity modulation of wind absorption
lines in an XRB and thus
providing a new measurement of its BH motion and orbital inclination angle.
We then apply this method to the HST/COS observations of
LMC~X--3 attempting to constrain the companion to primary mass ratio. Finally we discuss further
observations needed to achieve the required accuracy of BH mass estimate for LMC~X--3, as well as potential
problems and uncertainties of applying this method.

\section{Methodology}

So far, all BH masses in XRBs have been estimated using the Kepler's 3rd law of stellar motion, expressed in the
so called the mass function,
\begin{equation}
\label{equ:mass}
f(M)~\equiv~P_{\rm orb}K_{\rm C}^{3}/2\pi G~=~M_{\rm
BH}\sin^3i/(1+q)^{2},
\end{equation}
where $P_{\rm orb}$ is the orbital period, $K_{\rm C}$ is the semi--amplitude of the velocity curve of the
companion star, $M_{\rm BH}$ is BH mass, $i$ is the the orbital inclination angle, and $q~\equiv~M_{\rm
C}/M_{\rm BH}$ is the mass ratio. Since the only direct observables are $P_{\rm orb}$ and $K_{\rm C}$, both
$M_{\rm C}$ and $i$ have to be determined indirectly in order to obtain the BH mass estimate reliably. The
companion's mass $M_{\rm C}$ can be determined relatively reliably by the observed spectral type of the
companion star. For LMXBs, $i$ can be estimated by modeling the optical or infrared light curve modulation,
though model dependence and other uncertainties (such as accretion disk contamination) cannot be circumvented
completely. For HXMBs, $i$ is normally not determined very well; in many cases observations or lack of eclipse
of the accretion disk emission by the companion is used to put some constraints on the possible ranges of $i$.
For details of BH mass estimates using this method, please refer to \cite{2006ARA&A..44...49R}.

On the other hand, the mass ratio $q$ can be determined directly according to the law of momentum conservation,
i.e.,
\begin{equation}
\label{equ:ratio}
M_{\rm C}/M_{\rm BH}=K_{\rm BH}/K_{\rm C},
\end{equation}
if the semi--amplitude of the velocity curve of the BH $K_{\rm BH}$ can be observed directly. Since a BH is not
directly observable, we can only hope to observe any emission or absorption line feature co-moving with it. The
accretion disk certainly moves with the accreting BH. However any line feature of the inner accretion disk
suffers from the broadening of disk's orbital motion and distortions by relativistic effects around the BH, thus
making it practically impossible, or difficult to the least, for detecting the binary orbital motion of the BH.
Emission line features from the outer disk region are normally detected with double-peaks, which can be modeled
to obtain the semi--amplitude of the velocity of the compact object, as discussed above. However a major
uncertainty is to determine the mean separation between the emission regions of the blue-shifted and red-shifted
components.

Fortunately, as we have discussed above, absorption line features of accretion disk winds in BHXBs have been
routinely detected with high significance. This suggests that the accretion disk wind in an XRB moves with the
disk that produces the wind, since otherwise the wind from the disk would not intercept the continuum emissions
produced from the same disk. In this case, the Doppler motion of wind absorption line features can be considered
as that of the BH, unless the wind interacts strongly with the surrounding interstellar medium or the wind
intrinsic velocity has systematic orbital dependence. It has been found recently that large scale (pc) cavities
exist around microquasars (XRBs producing relativistic jets), and perhaps are ubiquitous in all BHXBs producing
strong winds (\citealt{2009ApJ...702.1648H}). This suggests that at least in the vicinity of the BH the wind
cannot interact directly with interstellar medium. The interactions of disk winds with interstellar medium at pc
scale would not complicate the observations, because it takes much longer than an orbital period before the
winds could arrive at the boundaries of these cavities, and thus the wind should have lost any possible memories
of the orbital motion of the compact object. For disk-fed accretion, the wind velocity is not expected to have
any orbital phase modulation; for stellar-wind accretion the focused wind may complicate the situation considerably
though. We therefore suggest to apply this method to only those systems in which the accretion is disk-fed,
i.e., no significant wind interaction happens between the stellar and accretion disk winds. Nevertheless only
observations can tell us in which kinds of systems this method can be used reliably.

\begin{figure}
\centerline{\includegraphics[width=0.8\textwidth, angle=0]{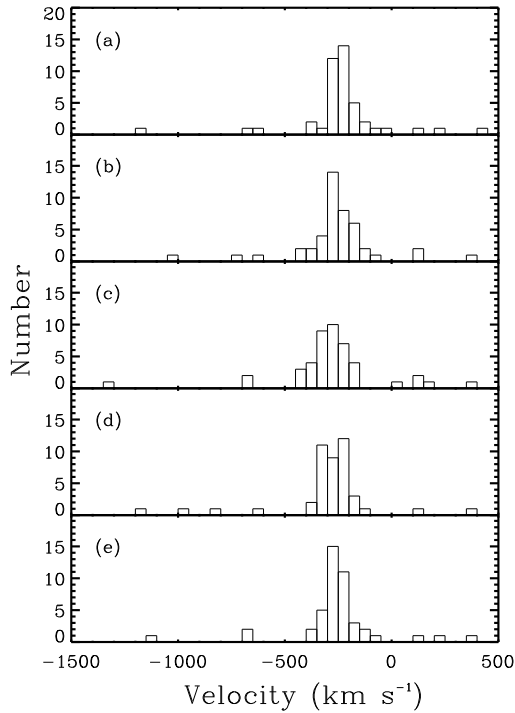}} \caption{Distributions of the line velocities
in four equally spaced intervals and their average values, during the Chandra/HETG observations lasting for more
than 60 ksec; note that here the system velocity of the BH XRB has already been subtracted. Panels from (a) to
(d) corresponding to the phases shown in Fig.~\ref{velocity} from left to right.
    \label{distr}
    }
\end{figure}

\begin{figure}
\centerline{\includegraphics[width=0.5\textwidth, angle=0]{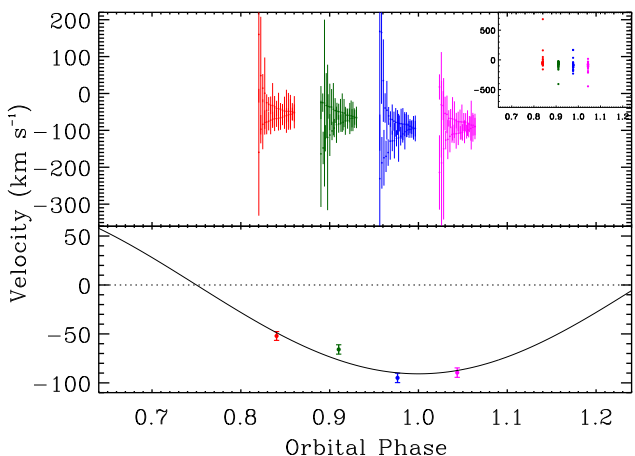}} \caption{Velocity curve of the 39 observed
absorption lines, after rejecting six lines as suspected outliers and subtracting the line of sight intrinsic
velocity of each line at each orbital phase. The upper panel marks the velocity of each line with its 1-$\sigma$
error bar slightly shifted horizontally for visual clarity; the inset shows all velocities, including the three
data points out of the range in the main panel. The bottom panel shows the weighted average velocity of all
lines in the upper panel at each phase; the solid curve is the fitted velocity curve with its orbital period and
phase fixed at the values observed previously.
    \label{velocity}
    }
\end{figure}

\section{Application to the LMXB GRO~J1655--40: a test study}

GRO~J1655--40, discovered as an X-ray transient by Zhang et al. (1994), is a well-known and best studied BHXB
and the second source with superluminal relativistic jets detected
(\citealt{1995Natur.374..141T,1995Natur.374..703H,1995Natur.375..464H}). Its system parameters remain so far the
best measured among all known BHXBs, with $P_{\rm orb}=2.62191\pm 0.00020$ d, $1/q~\equiv~M_{\rm BH}/M_{\rm
C}=2.6\pm 0.3$, $i=70.2\pm 1.9^\circ$, and $M_{\rm BH}=6.3\pm 0.5$~$M_\odot$ (all 95\% confidence)
(\citealt{1997ApJ...477..876O,2001ApJ...554.1290G}). Its precise BH mass and inclination measurements allow its
BH spin parameter determined from its X-ray continuum fitting, first proposed by Zhang et al. (1997) and then
refined by incorporating detailed modeling of various effects (e.g.,
\citealt{2005ApJ...619..446Y,2005ApJS..157..335L,2005ApJ...621..372D,2006ApJ...636L.113S,2009ApJ...701L..83S,2011CQGra..28k4009M}).
However possible sources of systematic errors are in the modeling of the ellipsoidal light modulation of the
companion star and contamination of the disk's continuum emission in the optical to infrared bands. Soria et al.
(1998) observed the velocity curve of the double-peaked disk emission line He~{\scriptsize II}~$\lambda$4686
from GRO~J1655--40, and derived its projected radial velocity semi-amplitude determined for the primary as
$K_{\rm BH}=76.2\pm 7.5$~km s$^{-1}$, yielding $M_{\rm BH}=6.62\pm 0.74$~$M_\odot$ and $i=66.6\pm 7.7^\circ$
(all 68.3\% confidence), fully consistent with that determined from the companion's velocity curve and
ellipsoidal light variation.

Miller et al. (2006) detected many narrow and ionized absorption lines with High Energy Transmission Grating
(HETG) Chandra observations of GRO~J1655--40. The exact position of the point source cannot be determined
accurately in the zeroth order image, which is severely piled-up due to its high flux. An offset in the source
position will cause a systematic shift of all lines in the wavelength space. However, this shift is mostly
canceled out in the combined total spectrum if each line is detected in the spectra of both the $+1$ and $-1$
orders with similar count spectra. However the combined lines will be broadened by the unknown systematic
offset, which in turn will cause larger uncertainties in determining the center of each line. There may also
exist other sources of systematic errors which can cause offset between the centers of the same line in the
spectra of the $+1$ and $-1$ orders.

Here we determine the position of the zeroth order image by minimizing the relative offsets between the
absorption lines in the $+1$ and $-1$ orders, and estimate the systematic error in determining the center
wavelength of each absorption line. We first fit each absorption line feature with a Gaussian profile in the
$+1$ and $-1$ orders of the High Energy Grating (HEG) and Medium Energy Grating (MEG) separately, i.e., we get a
pair of profiles for each absorption line, integrated over the whole observation in order to get the highest
signal to noise ratio. The local continuum around each line is assumed to have a power-law shape, but our
results are insensitive to the continuum shape. For HEG and MEG we obtain 43 and 9 pairs of absorption lines
with at least 3-$\sigma$ detection, resulting in 43 and 9 differences of the central wavelengths of these pairs,
respectively. We then use the Bayesian method to find the offset $d$ and its systematic error $\sigma_{\rm s}$
by maximizing the following function,
\begin{equation}p(d,\sigma_{\rm s})=\frac{1}{\sqrt{(2\pi)^{n}}}\prod\frac{1}{\sqrt{\sigma{_{x_{k}}}^2+\sigma_{\rm s}^2}}e^{-\frac{(d-x_k)^2}{2(\sigma{_{x_{k}}}^2+\sigma_{\rm s}^2)}},
\end{equation}
where $x_{k}$ and $\sigma{_{x_{k}}}$ are the wavelength difference and its statistical error of the $k$-th pair
($k=1$ to $n$, and $n=43$ for HEG or 9 for MEG), respectively. Finally for HEG we get
$d=(4.8\pm0.1)\times10^{-3}$ {\AA} and $\sigma_{\rm s}=(3.2\pm1.3)\times10^{-4}$ {\AA}.

We thus take the error for $d$ as $\sigma_{d}=\sigma_{\rm s}=3.2\times10^{-4}$ {\AA} for HEG. Similarly for MEG
we get $d=(9.7\pm0.8)\times10^{-3}$ {\AA} for MEG, where $\sigma_{d}=9\times10^{-4}$ {\AA} is dominated by the
systematic error. Taking these additional errors for MEG and HEG line wavelengths and requiring all pairs in HEG
with $d=4.8\times10^{-3}$ {\AA} and in MEG $d=9.7\times10^{-3}$ {\AA}, we have a total $\chi^2=55$ for 50
degrees of freedom, thus validating our method of determining the systematic errors. Since the wavelength bin
(an image pixel) is $2.5\times10^{-3}$ {\AA} or $5.0\times10^{-3}$ {\AA} for HEG or MEG, we thus shift the
spectra of the $-1$ and the $+1$ orders by one bin, towards the decreasing ($-1$ order) and increasing ($+1$
order) wavelength directions, respectively. This is identical to shifting the location of the point source by
one pixel in the zeroth order image. Finally we combine the two shifted spectra as one single spectrum.

Because the observations lasted for about a quarter of the orbital phase of GRO~J1655--40 and many absorption
lines have high signal to noise ratios, here we divide the observations into four equal orbital internals in
order to detect the orbital motion of the absorption lines. For each combined spectrum we again fit each
absorption feature with a Gaussian profile to determine its central wavelength, line width and intensity.
Table~1 lists the 45 lines in all four intervals with at least 3-$\sigma$ detection that are included for
further analysis; here only statistical errors are shown.

Fig.~\ref{distr} shows the velocity distributions of the four groups of 45 lines and their average velocities;
note that here the system velocity of the BH XRB has already been subtracted. Adopting the orbital period and
phase ephemeras from Greene et al. (2001) and assuming that the BH's motion is exactly anti-phased with the
companion's motion (Soria et al. 1998), we fit velocities of these lines simultaneously to a sinusoidal
function. In the fit, the intrinsic velocity of each line is a free parameter, but all lines in the same orbital
phase follow the same orbital modulation. This way we have 46 free parameters (45 intrinsic velocities for all
these lines plus the line of sight velocity amplitude of the black hole) with 180 data points. The fit results
in $K_{\rm BH}=93.8\pm 11.1~{\rm km~s^{-1}}$ with $\chi^2=179$ for 134 degrees of freedom; the systematic errors
determined above are included for all lines.

Considering that some lines may be outliers, we reject the four groups of six lines with central velocities more
than 300~${\rm km~s^{-1}}$ from the median values of each distribution, as listed in Table~1. The remaining four
groups of 39 lines have velocity dispersion (1-$\sigma$) of about 60~${\rm km~s^{-1}}$, i.e., the rejected lines
are more than 5-$\sigma$ away from the median values. The fit to the remaining four groups of 39 lines yields
$K_{\rm BH}=90.8\pm 11.3~{\rm km~s^{-1}}$ with $\chi^2=131$ for 116 degrees of freedom, a marginal improvement
over the full dataset fit. This means that the deviations of these possible outliers are not significant
statistically and thus do not deserve further studies at this stage. Fig.~\ref{velocity} shows the fitting
results; please note that the velocities are obtained by subtracting the fitted line of sight intrinsic velocity
of each line from the obtained central velocity of each line at this orbital phase. It is worth noting that the
phase zero in this system was defined as that when the companion is receding from the observer at the maximum
velocity (Orosz \& Bailyn 1997). Nevertheless both results are statistically consistent with $K_{\rm BH}=76.2\pm
7.5$~km s$^{-1}$ obtained by Soria et al. (1998). If we fit the four groups of 39 lines to a straight line, we
get $\chi^2=813$ for 154 degrees of freedom. The linear model is thus rejected with high significance, compared
to the sinusoidal model.

With the fitted $K_{\rm BH}=90.8\pm 11.3$~km s$^{-1}$ and taking the other system parameters (except the
inclination) of GRO~J1655--40 from Greene et al. (2001), we first obtain its BH mass $M_{\rm
BH}=5.41^{+0.98}_{-0.57}~M_{\odot}$ from Eq.~2 and then its system inclination $i=72.0^{+7.8}_{-7.5}~^\circ$
from Eq.~\ref{equ:mass}, respectively. All these (1-$\sigma$) errors are obtained by Monte-Carlo samplings,
because of the asymmetry and coupling of some errors. These parameters, albeit with large uncertainties due to
the very incomplete orbital coverage of observations, are consistent with all previous measurements. Therefore
the existing data of GRO~J1655--40 validate our proposal that absorption lines produced in the accretion disk
wind can be used to measure directly the orbital motion of the BHs in BHXBs.

\begin{figure}
   \centerline{
    \includegraphics[width=0.4\textwidth, angle=0]{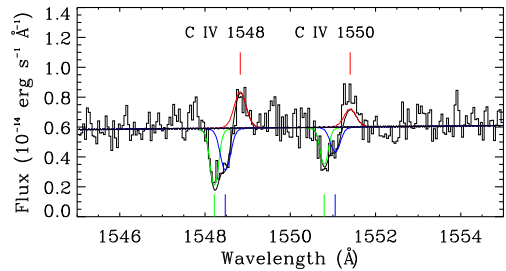}}
\caption{Spectrum of the available HST/COS observations around
     \civ\ doublet. Red curves indicate the fit to the emission from the
    heated surface of the companion star, green curves are the fit to
    the Galactic ISM absorptions at $0~{\rm km~s^{-1}}$ ,
    and blue curves indicate the fit to the accretion disk wind
    absorptions at $\approx50~{\rm km~s^{-1}}$
    (local velocity $\approx-400~{\rm km~s^{-1}}$). The vertical bars
    mark the central positions of these components.
     \label{fig:CIV}
}
\end{figure}

\begin{figure}
\centerline{
    \includegraphics[width=0.4\textwidth, angle=0]{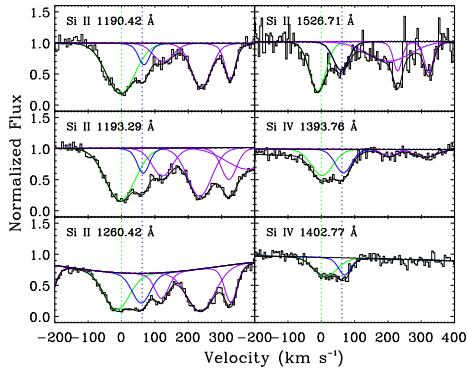}}
\caption{\siii\ and \siiv\
    absorption features in the spectrum of the previous HST/COS G130M observations, normalized to the continuum flux.
    The apparent higher velocity components (marked with pink curves)
    may be produced in the outer part of the disk wind at lower local
    velocities.
    \label{fig:si}
    }
\end{figure}

\section{Application to the HMXB LMC X--3}

LMC X--3 is another excellent object for applying this method. It is a bright XRB system in the Large Magellanic
Cloud composed of a B3~V star and a central BH, and receding away from us at a systemic velocity $V_{\rm
S}=+310~{\rm km~s^{-1}}$ (\citealt{cow83}; hereafter C83). It is one of the few BH systems that are persistently
luminous in both X-ray and far-ultraviolet (FUV) wavelength bands. The BH is believed to be undergoing accretion
from its B-star companion via Roche lobe overflow with an orbital period of 1.7 days (C83); we therefore do not
expect any significant contamination of the stellar wind to any absorption features of its accretion disk wind.
Spectroscopic observations of the B star indicate a large radial velocity semi-amplitude, $K_{\rm C}=235 \pm
11~{\rm km~s^{-1}}$ (C83). Taken the mass of the B3~V star as about 6~$M_\odot$, the BH mass in LMC X--3 is thus
estimated to be $5-10~M_\odot$, assuming an inclination angle of $50^\circ-70^\circ$ (C83; \citealt{kui88,
sor01}). Because of the considerable uncertainty in its inclination (and thus BH mass), its BH spin has not been
reliably determined yet with the X-ray continuum modeling method (e.g., \citealt{zha97, ste10}). Since the BH
should be moving at a comparable velocity to its companion, it is possible to directly measure its orbital
motion if the absorption lines produced in its accretion disk wind are detected.

Although LMC X--3 has been intensively studied in the X-ray band, the previously operated and currently
operating X-ray spectroscopic instruments lack the combination of the required sensitivity and resolution to
measure the expected Doppler motion of any accretion wind absorption features. High resolution UV spectroscopic
observations of this source are rather sparse; so far only two {\it FUSE} observations and four observations with
the Cosmic Origins Spectrograph (COS) onboard the {\it Hubble Space Telescope} (HST) (HST/COS hereafter) to LMC
X--3 have been made (\citealt{hut03, wang05, song10}, here after S10). Again {\it FUSE} does not have the
combination of the required sensitivity and the observations did not
produce spectra with high enough signal to noise ratio for the purpose of
this investigation. The HST/COS does have the required performance to do so,
although the available observations only covered a small portion of the
orbital period of LMC X--3. In this work, we
analyze the available HST/COS UV spectroscopic observations of LMC X--3, aiming
at constraining the systemic parameters and further demonstrating the
feasibility of our proposed new method of measuring the BH's mass in an
XRB.

From the {\it FUSE} and HST/COS observations, S10 detected variable \ovi\ and \nv\ emission lines and found that
variability of their intensities are inconsistent with expectation of a stellar wind origin. So they attributed
them to the heated stellar surface of the companion star. In fact both absorption and emission features are
detected in \nv\ (with G130M) and \civ\ (with G160M) doublets (Fig.~2 in S10) in the HST/COS observations
(Program 11642; please refer to S10 for detailed description of the HST/COS observations and data analysis).
However, since S10 focused on the origin of the emission variability and the connection between the emission
features and the systemic parameters, these absorption features have not been explored in detail.

Here we analyze the \civ\ doublet complexes (we do not conduct the same analysis to the \nv\ features because
the absorptions are much less significant than those of \civ). In Fig.~\ref{fig:CIV}, we decompose the observed
\civ\ doublet into three components for each line: one emission component and two absorption components. The
emission component is assumed to come from the heated surface of the companion, the same as the \ovi\ and \nv\
emission lines (S10); its width and Doppler shift are fixed to those inferred from the \ovi\ and \nv\ emission
lines in the decomposition. One absorption component (at zero velocity) is assumed to come from the Galactic ISM
absorption. The other significant absorption component has a redshift of about $50~{\rm km~s^{-1}}$, and may
have two possible origins. One is the absorption produced in the stellar wind of the B3~V companion star.
However, the orbital phase of the binary system was about 0.75 during the HST/COS observations, i.e., the
straight line connecting the companion star and the BH is perpendicular to our LOS (please be noted that, in
contrast to GRO~J1655--40, the orbital phase zero in LMC~X--3 was defined as that when the BH is at its superior
conjunction, i.e., the companion is just between the observer and the BH; Cowley 1983; S10).  In order for any
stellar wind to intercept significantly the emission from the inner disk region, a significant amount of stellar
wind has to stream to the inner disk region, i.e., the system must be wind-fed, against the common believe that
LMC X--3 is actually disk-fed. We thus rule out this possibility. The other and the only viable scenario is the
absorption by the accretion disk wind, making it possible to directly measure the orbital motion of the BH. It
is worth noting that the measured \civ\ column density in the ISM [$\log N_{\rm ISM}({\rm cm^{-2}})=13.7\pm0.1$]
agrees remarkably well with predicted value [$\log N_{\rm CIV}({\rm cm^{-2}})=13.8$] in a joint analysis of the
X-ray and FUV spectroscopic observations of LMC~X--3 (\citealt{yao09}), indirectly validating such a component
decomposition.

With the measured velocity $V_{0.75}=51.3\pm4.1~{\rm km~s^{-1}}$ of \civ\ absorption lines tracing the BH motion
and considering the systemic velocity ($V_{\rm S}=+310~{\rm km~s^{-1}}$) of LMC~X--3 and velocity semi-amplitude
of the companion star $K_{\rm C}=235~{\rm km~s^{-1}}$, we can constrain the companion to black hole mass ratio.
At phase 0.75, the BH is receding from us at the maximum speed, and thus the relation of $V_{0.75}$ and $K_{\rm
BH}$ can be expressed as
\begin{equation}
\label{equ:0.75} V_{0.75} = V_{\rm wind} + V_{\rm S} + K_{\rm BH},
\end{equation}
where $V_{\rm wind}$ is the intrisinc velocity of the accretion disk wind in our LOS. Assuming $V_{\rm
wind}=-400~{\rm km~s^{-1}}$, a similar wind velocity found in GRO~J1655--40 (Miller et al. 2006), from
Eq.~\ref{equ:0.75} we obtain $K_{\rm BH}\sim140~{\rm km~s^{-1}}$. Plugging these numbers in Eq.~\ref{equ:ratio},
we further obtain $q=M_C/M_{\rm BH}\sim0.6$, which is consistent with the ratio usually adopted for the LMC~X--3
system (e.g., C83). We therefore suggest that the observed \civ\ absorption feature is consistent to our model
that the accretion disk wind moves with the BH and the observed Doppler shift is a combination of the wind
velocity and the BH's orbital velocity in our LOS.

In the above exercise, we assumed a $V_{\rm wind}$ in order to obtain $K_{\rm BH}$, since there is a degeneracy
of $V_{\rm wind}$ and $K_{\rm BH}$ in Eq.~\ref{equ:0.75}. Unfortunately, the HST/COS G160M observation analyzed
above only lasted for one HST orbit, and thus cannot break the degeneracy, i.e., we cannot put an independent
measure of the BH's velocity, without observing the orbital modulation of the Doppler motion of the wind. The
HST/COS G130M observations in a shorter wavelength band covered a much larger part of the orbital period. As
shown in Fig.~\ref{fig:si}, the combined spectrum revealed several complicated absorption features.
Nevertheless, \siii\ and \siiv\ absorption lines at $V_{0.75}\approx 50~{\rm km~s^{-1}}$ are also detected; the
apparent higher velocity components may arise from the outer part of the accretion disk with lower local
velocities, mimicking the different velocities and ionization zones of AGN warm-absorbers/outflows (e.g.,
\citealt{arav05}). However, the combination of the complexity of the observed absorption features and the rather
incomplete orbital phase coverage of these previous HST/COS observations does not warrant further more
quantitative analysis for breaking the above mentioned degeneracy and probing the nature of those higher
velocity components.

Clearly, observations covering more orbital phases are badly needed to
break the degeneracy. Observations around phase 0.25 are the most
favorable ones for this purpose. In contrast to the existing observations
taken around 0.75 in which the BH is receding at the maximum velocity from
us, at phase 0.25, the BH is expected to be moving toward us at the maximum
velocity, so is the disk wind. Therefore, the relation between $V_{0.25}$
and $K_{\rm BH}$ can be expressed as
\begin{equation}
\label{equ:0.25} V_{0.25} = V_{\rm wind} + V_{\rm S} - K_{\rm BH}.
\end{equation}
If $V_{\rm S}=-400~{\rm km~s^{-1}}$ as assumed, the $V_{0.25}$ is expected
to be at $-230~{\rm km~s^{-1}}$. The real measurement of $V_{0.25}$ from
the future observations, together with $V_{0.75}$ measured from the
existing observations, would allow us to solve $V_{\rm wind}$ and
$K_{\rm BH}$ from Eqs.~\ref{equ:0.75} and \ref{equ:0.25} and then to
reliably constrain $M_C/M_{\rm BH}$ and system inclination angle $i$
(Eqs.~\ref{equ:ratio} and \ref{equ:mass}).

\section{Summary and Discussion}

As shown in Eq.~\ref{equ:ratio}, $M_{\rm BH}$ can be directly obtained if $K_{\rm BH}$ can be measured, in
addition to $M_{\rm C}$ and $K_{\rm C}$ measured by observing the companion star. In this work we suggest to
measure $K_{\rm BH}$ by detecting the Doppler orbital motion of the accretion disk wind absorption lines,
assuming that the accretion disk wind moves with the BH and does not have systematic orbital phase dependence.
This method has the potential of circumventing the model dependence and other uncertainties in estimating the
orbital inclination angle, that is required in the method commonly used to measure the BH masses in XRBs by
detecting the companion star's velocity and light curves, as shown in Eq.~\ref{equ:mass}. Actually knowing the
mass ratio with Eq.~\ref{equ:ratio}, one can in turn use Eq.~\ref{equ:mass} to derive the inclination angle,
which can be used to calibrate the light curve model used previously to derive the inclination angle.

Our analysis of the previous Chandra/HETG observations of GRO~J1655--40 have revealed wind velocity modulation
consistent with the orbital motion of the BH predicted from its previously measured system parameters. An
independent projected radial velocity semi-amplitude measured here allows its inclination angle determined
without using the modeling of its ellipsoidal light modulation of its companion. We find its BH radial velocity
semi-amplitude $K_{\rm BH}=90.8\pm 11.3$~km s$^{-1}$, BH mass $M_{\rm BH}=5.41^{+0.98}_{-0.57}~M_{\odot}$ and
system inclination $i=72.0^{+7.8}_{-7.5}~^\circ$ , where $M_{\rm BH}$ does not depend on $i$ at all. However the
very limited orbital coverage of the observations does not allow more accurate system parameter measurements of
this binary system. Nevertheless with the velocity component of its orbital motion removed, we can obtain more
accurate measurements of the intrinsic velocities of each line along our line of sight, and thus may be able to
constrain further the physical properties of the wind, by combining with the velocity broadenings of these
lines; this is the subject our future work.

Our analysis of the previous HST/COS observations of the HMXB LMC X--3 has
found absorption line features
consistent with that predicted by assuming the previously measured dynamical
parameters of LMC X--3 and the
wind properties in LMC X--3 being similar to that observed in another BHXB
GRO~J1655--40. Given the limitations of the previous HST/COS observations
of LMC X--3 that do not allow to break the degeneracy of the wind
velocity and the BH orbital velocity, new HST/COS observations are
required to cover significantly different orbital phases.

As mentioned in S10, the \civ\ features shown in Fig.~\ref{fig:CIV} might be P-cygni profiles. However, the
emission features agree well with all other emission lines detected, which are attributed to the heated stellar
surface. Therefore it is more reasonable to attribute the emission features in Fig.~\ref{fig:CIV} to the heated
stellar surface (as done in S10), thus invalidating the P-cygni profile interpretation. UV emission lines
produced from the heated stellar surface are not unique in the system of LMC~X--3, but rather a common feature
observed in XRBs (e.g., Vrtilek et al. 2003). Of course new observations suggested above would definitely reveal
the nature of the the \civ\ features shown in Fig.~\ref{fig:CIV}.

In this work, we have also assumed that the accretion disk wind velocity is constant, at least during one full
orbital phase, in order to apply this method reliably. In reality, the intrinsic wind velocity may have random
fluctuations, though the fluctuations do not seem to be significant in GRO~J1655--40 (\citealt{mil06a}).
However, it has been known that wind absorption features are not always detected and it is also not fully
understood when and why wind absorptions are present or absent. Future high signal to noise observations
covering more orbital phases may shed some lights on this problem and test ultimately if our suggested method
can be applied reliably and produce accurate BH mass measurements in BHXBs. The joint JAXA/NASA {\it ASTRO-H}
mission is particularly suitable for making such observations, with its high-throughput spectroscopy provided by
the micro-calorimeter with high spectral resolution of $\Delta E \sim7$ eV (\citealt{2010SPIE.7732E..27T}).
Finally we should point out that this method in principle can also be applied to other accreting compact
objects, such accreting neutron star and white dwarf binaries, with detectable accretion disk wind absorption
line features.

\section*{Acknowledgments}

S.N.Z. thanks the hospitality of CASA, University of Colorado at Boulder, during his visit in December 2010 when
the initial idea of this work emerged as results of some interesting discussions with local scientists there.
The anonymous referee is thanked for making several useful comments and suggestions. We appreciate comments and
suggestions made by Drs. Zhongxiang Wang and Hua Feng. Daniel Dewey is thanked for his information and
suggestion on evaluating the systematic errors. S.N.Z. also acknowledges partial funding support by the National
Natural Science Foundation of China under grant nos. 11133002, 10821061, 10725313, and by 973 Program of China
under grant 2009CB824800. Y.Y. appreciates financial support by NASA through grant HST-GO-11642.01-A.

\begin{table*}
\caption{Velocity and width of each absorption line at each orbital phase, detected with more than 3-$\sigma$
significance. For each orbital phase, the numbers in the left and right are the velocity shift and width
(broadening) of the absorption line; the 1-$\sigma$ errors are included in the parenthesis. The last column
indicates if the line is within 300~km s$^{-1}$ to the median velocity at each phase.}
\begin{tabular}{llccccc}
\hline
Ion and transition & Wavelength & Phase 1 & Phase 2 & Phase 3 & Phase 4 & Within 300\\
       & ($\AA$)    & (km s$^{-1}$) & (km s$^{-1}$) & (km s$^{-1}$) & (km s$^{-1}$) &(km s$^{-1}$) \\
\hline

Ni XXVI $1s^22s - 1s^25p$    & $6.1200$   & $-802(48)$   $102(88)$        & $-756(50)$   $115(106)$        & $-821(61)$   $152(90)$           & $-1338(94)$   $648(86)$       & $N$ \\
Ni XXVI $1s^22s - 1s^24p$    & $6.8163$   & $-535(33)$   $380(39)$        & $-560(29)$   $329(33)$          & $-478(34)$   $355(42)$           & $-471(33)$   $355(38)$          & $Y$ \\
Ni XXVI $1s^22s - 1s^23p$    & $9.0610$  & $-416(12)$   $212(15)$         & $-431(15)$   $241(20)$          & $-453(14)$ $215(18)$             & $-431(14)$   $220(18)$          & $Y$ \\
Ni XXVI $1s^22s - 1s^23p$    & $9.1050$  & $-432(16)$   $168(21)$         & $-432(13)$   $134(22)$          & $-465(16)$   $182(23)$           & $-445(18)$   $179(26)$          & $Y$ \\
Fe XXVI $1s - 2p$                     & $1.7798$   & $-1339(196)$ $861(239)$   & $-1187(246)$ $773(261)$     & $-1473(185)$ $962(207)$     & $-1107(141)$ $825(154)$      & $N$ \\
Fe XXV  $1s^2 - 1s2p$            & $1.8504$   & $ 74(142)$  $1408(157)$     & $-38(117)$   $1167(131)$     & $ -11(120)$  $2070(135)$      & $ 368(155)$   $1436(173)$    & $N$ \\
Fe XXIV $1s^22s - 1s^29p$   & $6.3475$   & $ 262(47)$   $289(57)$         & $ 250(38)$   $258(60)$          & $ 254(30)$   $103(91)$           & $ 253(30)$   $145(52)$           & $N$ \\
Fe XXIV $1s^22s - 1s^26p$   & $6.7870$   & $-413(11)$   $155(17)$        & $-410(14)$   $190(18)$          & $-425(11)$   $145(18)$           & $-426(12)$   $154(19)$           & $Y$ \\
Fe XXIV $1s^22s - 1s^25p$   & $7.1690$   & $-409(7)$     $261(9)$           & $-412(8)$      $255(9)$            & $-442(8)$    $244(9)$              & $-452(9)$      $280(10)$          & $Y$ \\
Fe XXIV $1s^22s - 1s^24p$    & $7.9893$  & $-344(8)$     $353(9)$          & $-357(9)$     $345(10)$           & $-390(9)$    $319(10)$             & $-405(8)$      $321(9)$            & $Y$ \\
Fe XXIV $1s^22s - 1s^23p$   & $10.6190$ & $-385(16)$   $295(21)$       & $-393(18)$   $318(23)$           & $-454(18)$   $286(23)$           & $-454(20)$    $335(23)$         & $Y$ \\
Fe XXIV $1s^22s - 1s^23p$   & $10.6630$ & $-358(19)$   $328(26)$        & $-388(19)$   $301(23)$          & $-404(19)$   $317(24)$           & $-386(19)$   $308(23)$          & $Y$ \\
Fe XXIII $2s^2 - 2s5p$             & $7.4722$  & $-337(16)$   $243(23)$         & $-351(16)$   $247(23)$         & $-330(16)$   $199(22)$            & $-359(16)$   $105(30)$          & $Y$ \\
Fe XXIII $2s^2 - 2s4p$             & $8.3029$  & $-204(13)$   $273(15)$         & $-224(14)$   $291(17)$         & $-296(14)$   $261(16)$            & $-280(15)$   $246(16)$          & $Y$ \\
Fe XXIII $2s^2 - 2s3p$             & $10.9810$  & $-379(16)$   $178(24)$       & $-387(24)$   $257(33)$         & $-398(17)$   $176(27)$           & $-390(16)$   $116(31)$          & $Y$ \\
Fe XXIII $2s^2 - 2s3p$             & $11.0180$  & $-333(18)$   $150(29)$       & $-315(19)$   $164(30)$         & $-360(19)$   $168(30)$           & $-355(19)$   $189(25)$          & $Y$ \\
Fe XXII  $2s^22p - 2s^23d$    & $11.7700$  & $-307(33)$   $147(54)$       & $-324(31)$   $112(67)$         & $-327(29)$   $95(51)$              & $-372(32)$   $127(61)$          & $Y$ \\
Fe XXII  $2s^22p - 2s^23d$   & $11.9200$  & $-286(50)$   $141(80)$       & $-281(62)$   $173(103)$        & $-358(56)$   $208(68)$           & $-324(57)$   $196(81)$          & $Y$ \\
Mn XXIV $1s^2 - 1s2p$           & $2.0062$   & $ 509(444)$  $1259(437)$  & $-582(261)$  $1018(322)$    & $- 7(302)$  $719(572)$           & $-326(184)$  $294(359)$       & $Y$ \\
Cr XXIV $1s - 2p$                     & $2.0901$   & $-303(156)$  $644(271)$    & $-262(180)$  $531(238)$      & $ 22(176)$   $404(334)$         & $-331(205)$  $735(209)$       & $Y$ \\
Cr XXIII $1s^2 - 1s2p$              & $2.1821$  & $-153(169)$  $852(220)$    & $-476(129)$  $341(353)$      & $-544(158)$  $732(225)$       & $-343(127)$  $497(169)$        & $Y$ \\
Ca XX $1s - 3p$                        & $2.5494$   & $-283(146)$  $364(183)$   & $-357(214)$  $442(294)$       & $-296(102)$  $345(160)$       & $-776(145)$  $544(182)$        & $Y$ \\
Ca XX $1s - 2p$                        & $3.0203$   & $-414(34)$   $393(48)$       & $-417(37)$   $422(45)$          & $-494(34)$   $388(45)$           & $-436(33)$   $356(49)$           & $Y$ \\
Ca XIX $1s^2 - 1s2p$              & $3.1772$   & $-391(48)$   $248(77)$       & $-395(50)$   $248(88)$          & $-547(58)$   $198(146)$         & $-427(46)$   $178(110)$         & $Y$ \\
Ar XVIII $1s - 2p$                      & $3.7329$   & $-431(25)$   $303(34)$       & $-383(32)$   $476(47)$          & $-442(31)$   $335(42)$           & $-468(26)$   $323(35)$            & $Y$ \\
S XVI $1s - 3p$                         & $3.9912$   & $-420(32)$   $256(50)$       & $-464(39)$   $242(58)$          & $-507(37)$   $197(72)$            & $-433(34)$   $219(59)$           & $Y$ \\
S XVI $1s - 2p$                         & $4.7292$   & $-421(36)$   $586(49)$       & $-520(35)$   $504(46)$          & $-533(25)$   $347(30)$            & $-355(23)$   $369(29)$           & $Y$ \\
S XV $1s2 - 1s2p$                   & $5.0387$   & $-359(32)$   $85(66)$          & $-348(41)$   $132(92)$          & $-415(40)$   $90(69)$              & $-526(96)$   $309(143)$         & $Y$ \\
Si XIV $1s - 3p$                        & $5.2172$   & $-360(30)$   $279(41)$       & $-437(31)$   $229(45)$          & $-545(34)$   $294(46)$           & $-470(23)$    $205(36)$           & $Y$ \\
Si XIV $1s - 2p$                        & $6.1822$   & $-446(10)$   $311(13)$       & $-464(11)$   $300(14)$          & $-483(11)$   $370(14)$           & $-453(10)$    $303(13)$           & $Y$ \\
Si XIII $1s^2 - 1s2p$                & $6.6480$   & $-343(29)$   $185(46)$       & $-335(32)$   $233(40)$          & $-375(25)$   $171(44)$           & $-373(33)$   $196(41)$            & $Y$ \\
Mg XII $1s - 7p$                        & $6.4486$   & $-408(27)$   $139(37)$      & $-395(21)$   $82(63)$            & $-412(34)$   $187(46)$            & $-411(28)$   $176(41)$            & $Y$ \\
Mg XII $1s - 5p$                        & $6.5801$   & $-510(17)$   $164(28)$      & $-502(16)$   $146(24)$          & $-510(17)$   $178(24)$            & $-521(20)$   $191(31)$            & $Y$ \\
Mg XII $1s - 3p$                        & $7.1062$   & $-365(16)$   $91(31)$        & $-406(17)$   $114(25)$          & $-398(20)$   $114(32)$            & $-381(18)$   $61(49)$              & $Y$ \\
Mg XII $1s - 2p$                        & $8.4210$  & $-410(9)$     $339(11)$       & $-427(9)$     $278(10)$           & $-467(9)$   $259(10)$              & $-471(10)$   $263(12)$            & $Y$ \\
Ne X $1s - 4p$                          & $9.7082$   & $-359(25)$   $151(46)$      & $-390(21)$   $107(39)$          & $-439(19)$ $95(30)$                 & $-399(21)$   $141(32)$           & $Y$ \\
Na XI $1s - 2p$                         & $10.0250$  & $-346(19)$   $166(26)$     & $-339(19)$   $146(25)$          & $-443(22)$   $151(27)$            & $-441(20)$   $127(26)$           & $Y$ \\
Ne X $1s−7p$                           & $9.2912$  & $-418(63)$   $84(93)$          & $-447(52)$   $93(84)$            & $-380(74)$   $199(63)$            & $-391(46)$   $100(80)$           & $Y$ \\
Ne X $1s - 6p$                          & $9.3616$   & $-776(25)$   $287(31)$      & $-846(25)$   $291(32)$          & $-837(34)$   $250(45)$            & $-945(32)$   $114(47)$            & $N$ \\
Ne X $1s - 5p$                          & $9.4807$   & $-365(28)$   $247(32)$      & $-322(28)$   $194(34)$           & $-370(27)$   $222(34)$           & $-391(26)$   $201(33)$           & $Y$ \\
Ne X $1s - 3p$                          & $10.2389$  & $-363(18)$   $222(25)$     & $-439(20)$   $229(22)$          & $-407(15)$   $143(21)$            & $-449(16)$   $207(20)$           & $Y$ \\
Ne X $1s - 2p$                         & $12.1330$   & $-441(21)$   $281(25)$     & $-420(21)$   $225(25)$          & $-452(20)$   $236(26)$            & $-482(17)$   $213(27)$           & $Y$ \\
Ne II                                            & $14.6310$   & $-389(34)$   $182(42)$    & $-393(33)$   $132(46)$           & $-410(30)$   $82(73)$               & $-457(46)$   $223(49)$         & $Y$ \\
O VIII $1s - 4p$                        & $15.1762$   & $-296(90)$   $361(113)$   & $-329(81)$   $314(133)$        & $-471(68)$   $186(101)$          & $-374(42)$   $41(73)$            & $Y$ \\
O VIII $1s - 3p$                        & $15.9870$   & $-23(63)$     $151(90)$     & $-9(77)$        $234(79)$           & $-110(74)$   $182(111)$          & $-2(80)$       $200(94)$          & $N$ \\
\hline
\end{tabular}
\end{table*}

\end{document}